\documentclass[slac_one]{revtex4}
\usepackage{graphicx}
\usepackage{fancyhdr}
\newcommand{\gsim}{\gtrsim}

\newcommand{\ord}[1]{\mathcal{O}{(#1)}}
\newcommand{\beq}{\begin{equation}}
\newcommand{\eeq}{\end{equation}}

\begin{document}

\title{The Little Randall-Sundrum Model at the LHC} 

%

\author{Hooman Davoudiasl}
\affiliation{Department of Physics, Brookhaven National Laboratory,
Upton, NY 11973, USA}

\begin{abstract}
We present a predictive warped model
of flavor, cut off at an ultraviolet scale $\ord{10^3}$~TeV, called 
the ``Little Randall-Sundrum (LRS)" model.  This model corresponds 
to a volume-truncation,
by a factor $y \approx 6$, of the RS scenario and is holographically
dual to dynamics with number of colors larger by $y$.  With separate gauge
and flavor dynamics, several unwanted contributions to precision
electroweak, $Z b\bar b$, and flavor observables are suppressed in
the LRS framework, compared with the corresponding RS case.  
The LRS truncation leads to a significant enhancement of the clean (golden)
di-lepton LHC signals, by $\ord{y^3}$.

\end{abstract}

\maketitle

\thispagestyle{fancy}



The following is based on a talk delivered by the author 
at ICHEP 2008, July 29-August 5, 2008, Philadelphia, PA,  USA.  
The source of material for this talk is Ref.~\cite{Davoudiasl:2008hx}, where more 
details and references can be found.  

The Randall-Sundrum (RS) model \cite{Randall:1999ee} 
was originally proposed to explain the enormous  
hierarchy between the inferred scale of gravity $M_P \sim 10^{19}$~GeV and 
the Standard Model (SM) weak scale of order 1~TeV.  
The RS model is based on a slice of 
warped AdS$_5$ geometry, bounded by UV (Planck) and IR (TeV) 
Minkowski branes.  The geometry provides an exponential redshift as one 
goes from the UV brane, characterized by a fundamental scale $M_5\sim M_P$, 
to the IR brane characterized by $e^{-k r_c \pi} M_5 \sim$~TeV, for 
$k r_c \pi \approx 35$.  Here, 
$k$ is the curvature scale and $r_c$ is the radius of compactification.  Thus, 
the hierarchy is generated exponentially, using natural parameters.  In the 
original model, all the SM fields were placed at the IR brane and the main signature 
of the model was a tower of TeV-scale spin-2  resonances, the 
Kaluza-Klein (KK)  gravitons \cite{Davoudiasl:1999jd}.

Later, gauge fields \cite{Davoudiasl:1999tf,Pomarol:1999ad} 
and fermions \cite{Grossman:1999ra} 
were placed in the 5D bulk, 
leading to an interesting model of flavor 
\cite{Gherghetta:2000qt} with TeV scale signatures.  Here, 
using 5D fermion masses, heavy fermions are localized towards 
the IR brane and light fermions are localized towards the UV brane.  However, 
a realistic model of flavor has very challenging signatures, as the light fermions, 
such as electrons, end up having small couplings to the KK modes. This 
suppresses production and clean di-lepton signals at colliders.  In addition,  
there is significant tension between precision data  
and TeV-scale warped flavor models.

Tree level oblique $S$ and $T$ corrections in RS models with bulk flavor, assuming no 
extra symmetry, are given by \cite{Agashe:2003zs}
\beq
S_{tree} \approx 2\pi\, (v/\kappa)^2\, 
\label{S2}
\eeq
and
\beq
T _{tree} \approx \frac{\pi}{2 \cos\theta_W^2}(v/\kappa)^2
(k r_c \pi)\, ,
\label{T2}
\eeq
where
$\kappa\equiv e^{-k r_c \pi} k$ is the KK scale  
and $\cos^2\theta_W\simeq 0.77$.  

Agreement with precision electroweak data requires $|S| \sim |T| \sim 0.1-0.3$.  
Taking $m_{KK} \approx 5$~TeV leads to $(S,T) \approx (0.1, 1.1)$, 
from the above formulas, in the RS model.  The large 
correction to $T_{tree}$  is from tree-level mixing of the gauge 
KK tower, induced by electroweak symmetry breaking (EWSB), enhanced 
by $k r_c \pi \approx 35$.

Here, one can then see that the truncation of the RS model by decreasing 
the value of $k r_c \pi$ can be helpful in reducing the tension between 
$T_{tree}$ and EW data, as this 
will reduce the strength of KK tower mixing caused by EWSB.  If 
$k r_c \pi = 6$ is assumed, one can still generate a hierarchy between 
a {\it flavor scale} $M_5 \sim 1000$~TeV and the weak scale, in a Little Randall-Sundrum 
(LRS) model \cite{Davoudiasl:2008hx}.  This flavor scale 
is large enough to address most if not all constraints from precision 
data.  Here, we define the truncation factor 
\beq
y \equiv
\frac{k r_c \pi|_{RS}}{k r_c \pi|_{LRS}}, 
\label{y}
\eeq
and  for our choice of scales $y\approx 6$.  It is assumed that 
the truncation in the LRS model leaves flavor physics unchanged, 
corresponding to keeping the 5D Yukawa coupling $\lambda_5$ and 
IR profiles of fermions at their RS values.

Equation (\ref{S2}) suggests that the LRS value of $S_{tree}$ does not change.  This 
is explained by noting that this quantity  is basically a result of universal shifts in 
light fermion couplings to gauge fields, induced by KK tower mixing of gauge fields, 
proportional to $\sqrt{k r_c \pi}$.  However, the light fermion coupling to gauge KK modes 
is propotional to $1/\sqrt{k r_c \pi}$.  As the net effect comes form a product of these factors, 
truncation leaves $S_{tree}$ unchanged. 

Note that there are also cutoff scale and  UV-sensitive loop contributions 
to $T$ that would push the KK masses to 10~TeV or more.  These contributions, as 
well as $T_{tree}$, 
can be eliminated with the assumption of a bulk 
$SU(2)_L \times SU(2)_R \times U(1)_X$ custodial symmetry \cite{Agashe:2003zs}, 
allowing $m_{KK} \gsim  3$~TeV, in both RS and LRS scenarios.

Next we will address non-oblique and flavor data, where 
more significant improvements can be obtained in the LRS 
framework.  By keeping the fermion IR-profiles at their RS value 
and $\lambda_5$ 
unchanged under truncation, 
the LRS model has the same level of flavor non-universality as its RS 
counterpart.  However, as the KK-mediated effects get truncated 
with reduced $k r_c \pi$, non-universal effects get suppressed within 
the LRS construct.  Two examples of 
these effects are contributions to 
$Z b {\bar b}$ coupling and $\Delta F = 2$ processes.  

Constraints on $Z b {\bar b}$ coupling
require non standard fermion representations under
the custodial symmetry, in addition to 
a $Z_2$ symmetry \cite{Agashe:2006at}, so that 
$m_{KK}^{RS} \sim 3$~TeV is allowed; otherwise,
$m_{KK}^{RS} \gsim 5$~TeV \cite{Agashe:2003zs}.
Without the custodial symmetry, a number of 
contributions to $Z b\bar b$ arise.
One is from the KK-tower mixing due to EWSB and the
enhanced coupling of the gauge KK modes to IR-localized $b_L$.
These corrections 
are proportional to $k r_c \pi$ and 
since the IR fermion profiles are kept fixed, 
get suppressed in the LRS scenario.  

The second type of correction to $Z b\bar b$ is due
to $\ord{1}$ mixing between $b_L$ and the exotic $SU(2)_R$
partner of $t_R$ \cite{KA}.  This contribution is absent 
if $t_R$ is in a representation that is a $SU(2)_{L,R}$
isosinglet \cite{Agashe:2006at}.  Note that without a bulk custodial symmetry,
there is no exotic
$t_R$ partner.  A third type of correction to $Z
b\bar b$ from the mixing of the KK modes of $b_R$ and the $b_L$ zero
mode is not truncated in the LRS model and proportional to $[1/m_{KK}(b_R)]^2$.  
Requiring corrections to $Z b\bar b$ coupling below $0.3\%$ then yields 
$m_{KK}(b_R)\gsim 4$~TeV.  However, we note that 
$m_{KK}\gsim 3$~TeV for gauge fields can accommodate 
this bound on the KK modes of
$b_R$, in the LRS framework, for a realistic set of fermion profiles 
\cite{Davoudiasl:2008hx}.  Therefore, 
all of the above constraints from $Z b\bar b$ can be satisfied for
gauge sector $m_{KK}\gsim 3$~TeV, without any custodial symmetries 
(however,  the $T$ parameter would still require protection).

We now consider the strongest constraints on generic bulk RS models,
from $\Delta F=2$ processes, due to tree level exchange of KK gluons.  The 
most stringent bound comes from excessive contributions 
$\delta(\epsilon_K)\propto k r_c \pi$ to $\epsilon_K$ from
$(V-A)\times (V+A)$ operators~\cite{Beall:1981ze,utfit}, 
requiring $m_{KK}^{RS}\gsim 20$~TeV; this bound is 
subject to roughly $30\%$ uncertainty \cite{Csaki:2008zd}.
In the LRS case, this contribution is
suppressed by $y$.  This suppression is likely not enough to allow 
for warped KK mode discovery at LHC and may require extra model building to 
bring the mass scales closer to the TeV regime.

{\it Phenomenology:} 
The LRS truncation leads to significant improvements in the LHC reach 
for the KK modes, because:
{\it (i)} Typically broad states~\cite{KKgluon} become narrower by
a factor $y\sim (0.2/0.08)^2$,
{\it (ii)} branching ratio (BR) into light fermions such as $e^+ e^-$ 
increases by a factor $y^2$, {\it (iii)} from {\it (i)} and
{\it (ii)} it follows that the signal ${\cal S}$ gets enhanced by $y^3 \sim 250$,
while the background ${\cal B}$ over the resonance width 
drops like $1/y$.
Hence, ${\cal S}/{\cal B}$ in the LRS model
is expected to go up by a remarkable factor of $y^4 \sim 1500$.

The enhanced discovery reach in the LRS model
could allow access
to the elusive EW gauge KK modes~\cite{Agashe:2007ki}.
For example, the $Z'\to \ell^+ \ell^-$, $\ell=e,\mu$,
golden decay modes which were quite challenging within the RS 
setup~\cite{Agashe:2007ki} could lead to discovery in the LRS model. 
Using the same
cuts as in Ref.~\cite{Agashe:2007ki}, a $Z^\prime$ with 
$M_{Z^\prime} =4-5~$TeV can be detectable with 100~fb$^{-1}$ in the LRS 
scenario; the reach within the RS model is $\sim 2$~TeV and requires 
1000~fb$^{-1}$ \cite{Agashe:2007ki}.

{\it Holography:} 
AdS/CFT correspondence
\cite{Maldacena:1997re} affords a dual description of the 
geometric RS results in terms of a 
strongly coupled large $N$
4D gauge theory.  The classical geometric relation
between the 4D gauge coupling $g_4$ and the 5D gauge
coupling $g_5$ \cite{Agashe:2002bx} is
\beq
1/g_4^2 = \tau_{\rm
UV} + \tau_{\rm IR} + \log(k/\kappa)/(k g_5^2), 
\label{g42}
\eeq
where, $\tau_{\rm UV}$ and $\tau_{\rm IR}$ will be treated as small
UV and IR quantum threshold corrections, respectively, and ignored.  
Keeping the value of $g_4$ fixed, reducing $k r_c \pi$ (the $\log$)
requires lowering the value of $k g_5^2$.  In the dual CFT,
this is interpreted as the contribution of
CFT ``quarks" to the running of external gauge couplings from the
fundamental scale, $M_5$, down to the TeV scale.  Then,  
$\sqrt{k g_5^2}\sim 4 \pi/\sqrt{N}$ should hold
between the dual theories. Thus, the LRS truncation 
is dual to a larger $N$ theory, $N^{LRS}\sim y N^{RS}>N^{RS}$, making the
inter-composite interactions weaker.  In particular, the Higgs-CFT (KK) 
interactions get weaker and lead to a smaller $T_{tree}$.

The main contribution
to $S_{tree}$ is from the universal vertex 
corrections~\cite{Agashe:2003zs}.  This is
governed by gauge
zero-KK mode mixing and scales as $1/\sqrt{N}$.  
The universal KK couplings to light
fermions, on the other hand, scales as $\sqrt{N}$. Therefore, $S$
remains the same under LRS truncation.

In our LRS construct, the 5D Yukawa coupling $\lambda_5$
is unscaled.  This is dual to
separate dynamics, characterized by a ``flavor" CFT
with $N_F\sim N^{RS}<N^{LRS}$.  

The non-oblique and FCNC contributions depend on the amount of
partial compositeness for a given $N_F$, in the dual picture.  The
amount of compositeness follows from the observed masses and mixing
angles \cite{aps}, once the value of $\lambda_5$ and 
the profile of $t_R$ are given.  The LRS partial 
compositeness is then unchanged by construction, and hence
the non-universal observables are suppressed by truncation, 
leading, in general, to better agreement with the
data.

In the LRS scenario, enhanced $\rho-photon$ mixing, 
proportional to $\sqrt N$ 
leads to larger couplings of light SM fermions
to composite modes.  The composite (KK) partial widths
into elementary fermions scales as $N$, however the
total width decreases as $1/N$.  Therefore, ${\cal S}\sim N^3$ and ${\cal B}\sim 1/N$,
over the resonance width.  These effects yield stronger LRS signals
at the LHC than for the RS counterpart, as $N^{LRS}/N^{RS} \gg 1$.

Here, we would like to note that other truncations motivated by other 
scales could still lead to improved discovery potential for warped models.  We chose 
$M_5\sim 1000$~TeV as a flavor scale, corresponding to a truncation 
factor $y\approx 6$.  However, one may choose to set the UV scale 
at, say, $10^{10}$~GeV, corresponding to the mass scale of right-handed 
neutrinos, in a seesaw scenario for neutrino masses.  Here, even for this 
relatively large UV scale, one still gets a 
significant enhancement of the light fermion signal, compared to the original RS case, since 
$y\approx 2$ and ${\cal S}\sim y^3$.  
Therefore, measuring the relative branching fractions of the light and heavy SM  
states in KK decays at the TeV scale, can potentially shed light on the 
size of the 5D slice, or the dual conformal window.

Even though the LRS model we presented here only addresses the flavor-weak 
hierarchy, a UV completion of this model can in principle accommodate the 
original Planck-weak hierarchy.  In fact, a recent model, based on a 6D geometry 
with 2 warped directions \cite{McDonald:2008ss} 
is a possible such completion.  In Ref.~\cite{McDonald:2008ss}, 
the LRS content resides on a  5D slice in a 6D space, where warping in one direction 
provides the redshift from $M_5\sim 1000$~TeV to 1~TeV, and warping along the 6$^{th}$ 
dimension provides the redshift from $M_6\sim M_P$ down to 1000~TeV.  Generalization 
to $n$-warped backgrounds, $n>2$, have also been discussed in Ref.~\cite{McDonald:2008ss}.

In summary, the LRS scenario offers a predictive  
framework to address flavor at a scale of order 1000~TeV, where warping  
generates the weak scale.  This model is a truncation of the original RS model.  
We assumed separate bulk gauge and flavor dynamics, leading 
to suppression of several unwanted contributions and lesser tension with 
precision data.  The LRS truncation leads to much improved prospects for discovery at the 
LHC in the dilepton channel, 
compared to its RS counterpart.  Given the sensitivity of collider phenomenology 
to  truncation, one may use TeV-scale data to probe the size of the 5D bulk or 
the dual conformal window.  The LRS model may be UV completed to account 
for the Planck-weak hierarchy.  An example of such completion has been proposed 
in Ref.~\cite{McDonald:2008ss}.

\begin{acknowledgments}
Work supported by Department of Energy contract DE-AC02-98CH10886.
\end{acknowledgments}

\end{document}